# A proposal for the optimal estimation of states in Quantum Information Processing

**Mario Mastriani**

CIT-DGIIT-ANSES, 353 Piedras St., 2nd. Floor, Room 228 (C1070AAG), Buenos Aires, Argentina.
mmastri@anses.gov.ar

*Abstract*—An optimal estimator of quantum states based on a modified Kalman's Filter is proposed in this work. Such estimator acts after state measurement, allowing obtain an optimal estimation of quantum state resulting in the output of any quantum algorithm**.** This method is much more accurate than other types of quantum measurements, such as, weak measurement, strong measurement, quantum state tomography, among others.

## 1  Introduction

The Problem of Measurement in quantum mechanics [1] has been defined in various ways, originally by scientists, and more recently by philosophers of science who question the foundations of quantum mechanics. Measurements are described with diverse concepts in quantum physics such as:

- wave functions (probability amplitudes) evolving unitarily and deterministically (preserving information) according to the linear Schrödinger equation,
- superposition of states, i.e., linear combinations of wave functions with complex coefficients that carry phase information and produce interference effects (the principle of superposition),
- quantum jumps between states accompanied by the "collapse" of the wave function that can destroy or create information (Dirac's projection postulate, von Neumann's Process),
- probabilities of collapses and jumps given by the square of the absolute value of the wave function for a given state,
- values for possible measurements given by the eigenvalues associated with the eigenstates of the combined measuring apparatus and measured system (the axiom of measurement),
- the Heisenberg indeterminacy principle.

The original problem, said to be a consequence of Niels Bohr's "Copenhagen interpretation" of quantum mechanics, was to explain how our measuring instruments, which are usually macroscopic objects and treatable with classical physics, can give us information about the microscopic world of atoms and subatomic particles like electrons and photons.

Bohr's idea of "complementarity" insisted that a specific experiment could reveal only partial information - for example, a particle's position. "Exhaustive" information requires complementary experiments, for example to determine a particle's momentum (within the limits of Werner Heisenberg's indeterminacy principle).

Some define the problem of measurement simply as the logical contradiction between two laws describing the motion of quantum systems; the unitary, continuous, and deterministic time evolution of the Schrödinger equation versus the non-unitary, discontinuous, and indeterministic collapse of the wave function. John von Neumann saw a problem with two distinct (indeed, opposing) processes.

The mathematical formalism of quantum mechanics provides no way to predict when the wave function stops evolving in a unitary fashion and collapses. Experimentally and practically, however, we can say that this occurs when the microscopic system interacts with a measuring apparatus.

Others define the measurement problem as the failure to observe macroscopic superpositions.

Decoherence theorists (e.g., H. Dieter Zeh and Wojciech Zurek, who use various non-standard interpretations of quantum mechanics that deny the projection postulate - quantum jumps - and even the existence of particles), define the measurement problem as the failure to observe superpositions such as Schrödinger's Cat. Unitary time evolution of the wave function according to the Schrödinger wave equation should produce such macroscopic superpositions, they claim.

Information physics treats a measuring apparatus quantum mechanically by describing parts of it as in a metastable state like the excited states of an atom, the critically poised electrical potential energy in the discharge tube of a Geiger counter, or the supersaturated water and alcohol molecules of a Wilson cloud chamber. (The pi-bond orbital rotation from cis- to trans- in the light-sensitive retinal molecule is an example of a critically poised apparatus).

Excited (metastable) states are poised to collapse when an electron (or photon) collides with the sensitive detector elements in the apparatus. This collapse is macroscopic and irreversible, generally a cascade of quantum events that release large amounts of energy, increasing the (Boltzmann) entropy. But in a "measurement" there is also a local decrease in the entropy (negative entropy or information). The global entropy increase is normally orders of magnitude more than the small local decrease in entropy (an increase in stable information or Shannon entropy) that constitutes the "measured" experimental data available to human observers.

The creation of new information in a measurement thus follows the same two core processes of all information creation - quantum cooperative phenomena and thermodynamics. These two are involved in the formation of microscopic objects like atoms and molecules, as well as macroscopic objects like galaxies, stars, and planets.

According to the correspondence principle, all the laws of quantum physics asymptotically approach the laws of classical physics in the limit of large quantum numbers and large numbers of particles. Quantum mechanics can be used to describe large macroscopic systems.

Does this mean that the positions and momenta of macroscopic objects are uncertain? Yes, it does, although the uncertainty becomes vanishingly small for large objects, it is not zero. Niels Bohr used the uncertainty of macroscopic objects to defeat Albert Einstein's several objections to quantum mechanics at the 1927 Solvay conferences.

But Bohr and Heisenberg also insisted that a measuring apparatus must be a regarded as a purely classical system. They can't have it both ways. Can the macroscopic apparatus also be treated by quantum physics or not? Can it be described by the Schrödinger equation? Can it be regarded as in a superposition of states?

The most famous examples of macroscopic superposition are perhaps Schrödinger's Cat, which is claimed to be in a superposition of live and dead cats, and the Einstein-Podolsky-Rosen experiment, in which entangled electrons or photons are in a superposition of two-particle states that collapse over macroscopic distances to exhibit properties "nonlocally" at speeds faster than the speed of light.

These treatments of macroscopic systems with quantum mechanics were intended to expose inconsistencies and incompleteness in quantum theory. The critics hoped to restore determinism and "local reality" to physics. They resulted in some strange and extremely popular "mysteries" about "quantum reality," such as the "many-worlds" interpretation, "hidden variables," and signaling faster than the speed of light.

We develop a quantum-mechanical treatment of macroscopic systems, especially a measuring apparatus, to show how it can create new information. If the apparatus were describable only by classical deterministic laws, no new information could come into existence. The apparatus need only be adequately determined, that is to say, "classical" to a sufficient degree of accuracy.

Everything said so far indicates the sensible which is the performance of quantum computing to the correct measurement of the quantum states.

On the other hand, a new technology allows us to avoid the problem of quantum measurement [2, 3]. However, this technology allows work exclusively with Computational Basis States (CBS), i.e., pure and orthogonal quantum base states.

Therefore, a new method of quantum measurement in the case of generic qubits becomes imperative (i.e., not just for CBS) and more accurate than the methods currently in use [4-27]. Thus, in this work, we present a novel proposal to recover quantum state to the output of a quantum algorithm after its measurement via a modified Kalman's Filter [28-32], and Recursive Least Squares (RLS) filter [33-35], too. This is the essence of this work, which is organized as follows:

Preliminaries to new quantum measurement method are outlined in Sect. 2. A tour from Schrodinger equation to quantum algorithms is presented in Sect. 3. The new method (optimal state estimator) is outlined in Sect. 4. Finally, Sect. 5 provides a conclusion and future works proposal of the paper.

## 2 Preliminaries to new quantum measurement method

In this section, we present the following topics:

- Wave function collapse
- Quantum Measurement Problems
- Before and after measurement
- Types of measurement and state reconstruction

### 2.1 Wave function collapse

In quantum mechanics, wave function collapse is the phenomenon in which a wave function -initially in a superposition of several eigenstates- appears to reduce to a single eigenstate after interaction with a measuring apparatus [36]. It is the essence of measurement in quantum mechanics, and connects the wave function with classical observables like position and momentum. Collapse is one of two processes by which quantum systems evolve in time; the other is continuous evolution via the Schrödinger equation [37]. However in this role, collapse is merely a black box for thermodynamically irreversible interaction with a classical environment [38]. Calculations of quantum decoherence predict apparent wave function collapse when a superposition forms between the quantum system's states and the environment's states. Significantly, the combined wave function of the system and environment continue to obey the Schrödinger equation [39].

When the Copenhagen interpretation was first expressed, Niels Bohr postulated wave function collapse to cut the quantum world from the classical [40]. This tactical move allowed quantum theory to develop without distractions from interpretational worries. Nevertheless it was debated, for if collapse were a fundamental physical phenomenon, rather than just the epiphenomenon of some other process, it would mean nature were fundamentally stochastic, i.e. nondeterministic, an undesirable property for a theory [38, 41]. This issue remained until quantum decoherence entered mainstream opinion after its reformulation in the 1980s [38, 39, 42]. Decoherence explains the perception of wave function collapse in terms of interacting large- and small-scale quantum systems, and is commonly taught at the graduate level (e.g. the Cohen-Tannoudji textbook) [43]. The quantum filtering approach [44-47] and the introduction of quantum causality non-demolition principle [48] allows for a classical-environment derivation of wave function collapse from the stochastic Schrödinger equation.

## 2.2 Quantum Measurement Problems

The measurement problem in quantum mechanics is the unresolved problem of how (or if) wave function collapse occurs. The inability to observe this process directly has given rise to different interpretations of quantum mechanics, and poses a key set of questions that each interpretation must answer. The wave function in quantum mechanics evolves deterministically according to the Schrödinger equation as a linear superposition of different states, but actual measurements always find the physical system in a definite state. Any future evolution is based on the state the system was discovered to be in when the measurement was made, meaning that the measurement "did something" to the process under examination. Whatever that "something" may be does not appear to be explained by the basic theory.

To express matters differently (to paraphrase Steven Weinberg [4, 5]), the Schrödinger wave equation determines the wave function at any later time. If observers and their measuring apparatus are themselves described by a deterministic wave function, why can we not predict precise results for measurements, but only probabilities? As a general question: How can one establish a correspondence between quantum and classical reality? [6].

## 2.3 Before and after measurement

In quantum mechanics, measurement is a non-trivial and highly counter-intuitive process. Firstly, because measurement outcomes are inherently probabilistic, i.e. regardless of the carefulness in the preparation of a measurement procedure, the possible outcomes of such measurement will be distributed according to a certain probability distribution. Secondly, once a measurement has been performed, a quantum system in unavoidably altered due to the interaction with the measurement apparatus. Consequently, for an arbitrary quantum system, pre-measurement and post-measurement quantum states are different in general [49].

***Postulate.*** Quantum measurements are described by a set of measurement operators $\{\hat{M}_m\}$, index $m$ labels the different measurement outcomes, which act on the state space of the system being measured. Measurement outcomes correspond to values of *observables*, such as position, energy and momentum, which are Hermitian operators [49, 50] corresponding to physically measurable quantities.

Let $|\psi\rangle$ be the state of the quantum system immediately before the measurement. Then, the probability that result $m$ occurs is given by

$$p(m) = \langle\psi|\hat{M}_m^\dagger \hat{M}_m|\psi\rangle \tag{1}$$

and the post-measurement quantum state is

$$|\psi\rangle_{pm} = \frac{\hat{M}_m|\psi\rangle}{\sqrt{\langle\psi|\hat{M}_m^\dagger \hat{M}_m|\psi\rangle}} \tag{2}$$

Operators $\hat{M}_m$ must satisfy the completeness relation, i.e., $\sum_m \hat{M}_m^\dagger \hat{M}_m = I$ [49] because that guarantees that probabilities will sum to one: $\sum_m \langle\psi|\hat{M}_m^\dagger \hat{M}_m|\psi\rangle = \sum_m p(m) = 1$.

Let us work out a simple example. Assume we have a polarized photon with associated polarization orientations 'horizontal' and 'vertical'. The horizontal polarization direction is denoted by $|0\rangle$ and the vertical polarization direction is denoted by $|1\rangle$. Thus, an arbitrary initial state for our photon can be described by the quantum state $|\psi\rangle = \alpha|0\rangle + \beta|1\rangle$, where $\alpha$ and $\beta$ are complex numbers constrained by the normalization condition $|\alpha|^2 + |\beta|^2 = 1$ and $\{|0\rangle, |1\rangle\}$ is the computational basis spanning $H^2$.

Now, we construct two measurement operators $\hat{M}_0 = |0\rangle\langle 0|$ and $\hat{M}_1 = |1\rangle\langle 1|$ and two measurement outcomes $a_0$, $a_1$. Then, the full observable used for measurement in this experiment is $\hat{M} = a_0|0\rangle\langle 0| + a_1|1\rangle\langle 1|$. According to Postulate, the probabilities of obtaining outcome $a_0$ or outcome $a_1$ are given by $p(a_0) = |\alpha|^2$ and $p(a_1) = |\beta|^2$. Corresponding post-measurement quantum states are as follows: if outcome = $a_0$ then $|\psi\rangle_{pm} = |0\rangle$; if outcome = $a_1$ then $|\psi\rangle_{pm} = |1\rangle$.

2.4 Types of measurement and state reconstruction

As we have seen in the previous subsection, quantum measurement is not a minor issue [4-6]. In fact, it is an issue still unresolved [7, 8], which would make it impossible for every practical effort to implement any genuine quantum algorithm in general and quantum image processing algorithm in particular. Really, it is an inherited problem of quantum physics and known as the paradox of measurement [9-12].

From a practical point of view, inside context of quantum image processing, the problem is reduced to the following: suppose we develop a quantum algorithm for filtering classic images. A first problem would be (no doubt), how to introduce a classical noisy image within the quantum computer? That is to say, design of the interfaces (classical-to-quantum, and quantum-to classical). But, the second would be, how to measure the results of a quantum filtering algorithm, and to take the result of that filtering process and carry out to the classical world, in other words, the recovery of the classical version of the filtered image into its original space, i.e., the classic world where it was generated. It is obvious that an absolutely accurate technique of measurement is needed. Unfortunately, all efforts in this regard have been useless [13, 14].

However, in the last decade there have been several efforts to remedy this situation, namely:

- Weak measurement
- Restoring the quantum state
- Quantum state tomography

**Weak measurement** is a technique to measure the average value of a quantum *observable* $|\psi\rangle_{pm}$ without appreciably affecting the initial state $|\psi\rangle$ of the system being measured [15-19]. Weak measurements differ from normal (sometimes called "strong" or "von Neumann") measurements in two ways:

1. If $|\psi\rangle_{pm}$ has discrete spectrum (which we assume for simplicity), a strong measurement when the system is in state $|\psi\rangle$ yields an eigenvalue of $|\psi\rangle_{pm}$; if the measurement is repeated many times (starting each time with the system in state $|\psi\rangle$) one obtains a sequence of eigenvalues of $|\psi\rangle_{pm}$ which when averaged yield an approximation to $\langle\psi|\psi_{pm}|\psi\rangle$, the expectation of $|\psi\rangle_{pm}$ in the state $|\psi\rangle$.

By contrast, a *weak measurement* only yields a sequence of numbers which average to $\langle\psi|\psi_{pm}|\psi\rangle$. For example, a strong measurement of the spin of a spin-1/2 particle must yield spin 1/2 or -1/2, but a particular weak measurement could yield spin 100, while a subsequent weak measurement on an identical system might be -128.3 . Typically, a single weak measurement gives little information; only the average of a large number of such measurements is meaningful.

2. A strong measurement changes ("projects") an initial pure state $|\psi\rangle$ to an eigenvector of $|\psi\rangle_{pm}$. (The particular eigenvector obtained cannot be predicted, though its probability is determined.) This substantially changes the state $|\psi\rangle$ unless $|\psi\rangle$ happened to be close to that eigenvector.

However, a weak measurement does not substantially change the initial state.

Weak measurements are usually implemented by coupling the original system $\Psi$ to be measured with an auxiliary quantum "meter system" $M$. The meter along a scale, though in practice various microscopic quantum systems are used. The composite system is mathematically represented as the tensor product of $\Psi$ with $M$, denoted $\Psi \otimes M$. A "product" state in this tensor product is typically denoted $|\psi\rangle|m\rangle$, where $|\psi\rangle$ is a state of $\Psi$ and $|m\rangle$ a state of $M$. States which are not product states are called *entangled* states.

The results obtained by this technique are as weak as its name, therefore, we proceed to the next.

**Restoring the quantum state** is an effort to recover the original state $|\psi\rangle$ from the alleged invertibility of measurement operator through the matrix that represents, that is to say $\hat{M}$ of Subsect. 2.3 [20]. Parrott work is presented in opposition to the technique of weak measurement in general and Katz et al work [21] in particular. Other relevant works mediate between the above [22, 23], also without success.

Today, we know based on Stochastic Processes and Adaptive Filtering [28-35] the single matrix inversion in the process of estimation or identification does not restore the state of a system hidden behind such matrix. This is due to the need to model correctly state and measurement noises and the appropriate architecture of the estimator for the correct system state recovery from the observables. This deficiency explains why Wiener filter was completely replaced by the Kalman's filter in the presence of said noise [28-32]. Therefore, this technique is as weak as that at which it opposes.

**Quantum state tomography** is the process of reconstructing the quantum state (density matrix) for a source of quantum systems by measurements on the systems coming from the source [24, 25]. Being the density matrix for pure or mixed states,

$$\hat{\rho} = \sum_m p(m) |\psi_m\rangle\langle\psi_m| \qquad (3)$$

The source may be any device or system which prepares quantum states either consistently into quantum pure states or otherwise into general mixed states. To be able to uniquely identify the state, the measurements must be tomographically complete. That is, the measured operators must form an operator basis on the Hilbert space of the system, providing all the information about the state. Such a set of observations is sometimes called a quorum. In quantum process tomography on the other hand, known quantum states are used to probe a quantum process to find out how the process can be described. Similarly, quantum measurement tomography works to find out what measurement is being performed. The general principle behind quantum state tomography is that by repeatedly performing many different measurements on quantum systems described by identical density matrices, frequency counts can be used to infer probabilities, and these probabilities are combined with Born's rule to determine a density matrix which fits the best with the observations [26, 27]. Obviously, this method is a spartan estimator of the density matrix and not the states themselves. In fact, it is a monitor of the elements of the matrix, only. Therefore, our problem persists.

## 3 From Schrodinger equation to quantum algorithms

3.1 Schrödinger´s equation and unitary operators

A quantum state can be transformed into another state by a unitary operator, symbolized as $U$, with $U^\dagger U = I$ (where $I$ is the identity matrix), which is required to preserve inner products: If we transform $|\chi\rangle$ and $|\psi\rangle$ to $U|\chi\rangle$ and $U|\psi\rangle$, then $\langle\chi|U^\dagger U|\psi\rangle = \langle\chi|\psi\rangle$, and being $|\chi\rangle$ and $|\psi\rangle$ two wave functions. In particular, unitary operators preserve lengths: $\langle\psi|U^\dagger U|\psi\rangle = \langle\psi|\psi\rangle = 1$.

On the other hand, the unitary operator satisfies the following differential equation known as the Schrödinger equation [50-53]:

$$\frac{d}{dt}U(t) = \frac{-i\hat{H}}{\hbar}U(t) \tag{4}$$

where $\hat{H}$ represents the Hamiltonian matrix of the Schrödinger equation, $i = \sqrt[2]{-1}$, and $\hbar$ is the Planck constant. Multiplying both sides of Eq. 4 by $|\psi(0)\rangle$ and setting $|\psi(t)\rangle = U(t)|\psi(0)\rangle$ yields

$$\frac{d}{dt}|\psi(t)\rangle = \frac{-i\hat{H}}{\hbar}|\psi(t)\rangle \tag{5}$$

The solution to the Schrödinger equation is given by the matrix exponential of the Hamiltonian matrix (time invariant):

$$U(t) = e^{\frac{-i\hat{H}t}{\hbar}} \tag{6}$$

Thus the probability amplitudes evolve across time according to the following equation:

$$|\psi(t)\rangle = e^{\frac{-i\hat{H}t}{\hbar}}|\psi(0)\rangle \tag{7}$$

Equation 7 is the main piece in building circuits, gates and quantum algorithms, being $U$ who represents such elements [50].

Finally, the discrete version of Eq. 5 is

$$|\psi_{t+1}\rangle = \frac{-i\hat{H}}{\hbar}|\psi_t\rangle \tag{8}$$

Equation 8 is the foundation on which we build the optimal estimator of quantum states.

3.2 Quantum Circuits, Gates and Algorithms

As we can see in Fig. 1, and remember Eq. 8, the quantum algorithm (identical case to circuits and gates) viewed as a transfer (or mapping input-to-output) has two types of output:

a) the result of algorithm (circuit of gate), i.e., $|\psi_{t+1}\rangle$
b) part of the input $|\psi_t\rangle$, i.e., $|\underline{\psi}_t\rangle$ (underlined $|\psi_t\rangle$), in order to impart reversibility to the circuit, which is a critical need in quantum computing [1].

Besides, we can see clearly a module for measuring $|\psi_{t+1}\rangle$ (which will be extensively discussed in the next section) with their respective output, i.e., $|\varphi_{t+1}\rangle$, and a number of elements needed for the physical implementation of the quantum algorithm (circuit or gate), namely: control, ancilla and trash [50]. In this figure as well as in the rest of them (unlike [50]) a single fine line represents a wire carrying $1$ qubit or $N$ qubits (qudit), interchangeably, while a single thick line represents a wire carrying *1* or *N* classical bits, interchangeably too.

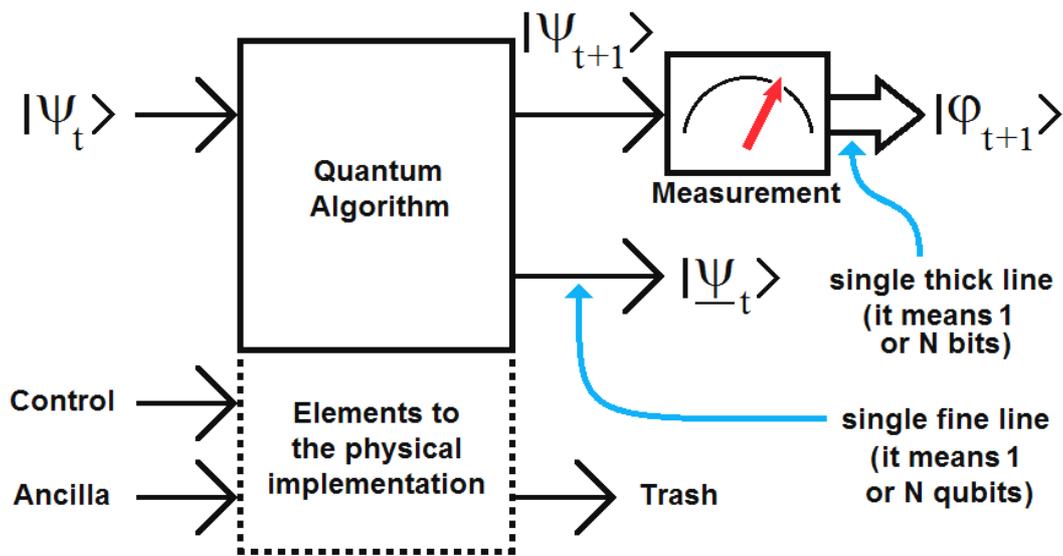

**Fig. 1** Module to measuring, quantum algorithm and the elements needs to its physical implementation.

However, the mentioned concept of reversibility is closely related to energy consumption, and hence to the Landauer's Principle.

On the other hand, computational complexity studies the amount of time and space required to solve a computational problem. Another important computational resource is energy. In this section, we study the energy requirements for computation. Surprisingly, it turns out that computation, both classical and quantum, can in principle be done without expending any energy! Energy consumption in computation turns out to be deeply linked to the reversibility of the computation.

What is the connection between energy consumption and irreversibility in computation? Landauer's principle provides the connection, stating that, in order to erase information, it is necessary to dissipate energy. More precisely, Landauer's principle may be stated as follows:

<u>Landauer's principle (first form):</u> *Suppose a computer erases a single bit of information. The amount of energy dissipated into the environment is at least $k_B T \ln 2$, where $k_B$ is a universal constant known as Boltzmann's constant, and T is the temperature of the environment of the computer.*
According to the laws of thermodynamics, Landauer's principle can be given an alternative form stated not in terms of energy dissipation, but rather in terms of entropy:

<u>Landauer's principle (second form):</u> *Suppose a computer erases a single bit of information. The entropy of the environment increases by at least $k_B \ln 2$, where $k_B$ is Boltzmann's constant.*

Consider a gate like the gate, which takes as input two bits, and produces a single bit as output. This gate is intrinsically irreversible because, given the output of the gate, the input is not uniquely determined. For example, if the output of the gate is 1, then the input could have been any one of 00, 01, or 10. On the other hand, the gate is an example of a reversible logic gate because, given the output of the gate, it is possible to infer what the input must have been. Another way of understanding irreversibility is to think of it in terms of information erasure. If a logic gate is irreversible, then some of the information input to the gate is lost irretrievably when the gate operates – that is, some of the information has been erased by the gate. Conversely, in a reversible computation, no information is ever erased, because the input can always be recovered from the output. Thus, saying that a computation is reversible is equivalent to saying that no information is erased during the computation.

Summing-up, the above expressed justifies the inexcusable need for the presence of $|\underline{\psi}_t\rangle$ to the output of quantum gate [50].

## 4 Optimal State Estimator (OSE)

### 4.1 Classical state estimator in noiseless environments

In order to develop an optimal estimate of quantum states, we start defining everything on a classical type of estimator called Recursive Least Squre RLS [33-35] and derived from the famous Kalman's filter [28-32]. Such estimator (time discrete version and in noiseless environment) is based on Fig. 2, in which,

A: plant $\in \mathbb{R}^{N \times N}$
M: measurement operator $\in \mathbb{R}^{M \times N}$
$\Delta$: unitary delay $(N \times N)$
t: time
X: state to be estimated $\in \mathbb{R}^{N \times 1}$
Y: observable $\in \mathbb{R}^{M \times 1}$
$\varepsilon$: error of estimation $\in \mathbb{R}^{M \times 1}$
K: Kalman's gain $\in \mathbb{R}^{N \times M}$
$\hat{X}$ : estimated state $\in \mathbb{R}^{N \times 1}$
$\hat{Y}$ : output of estimator $\in \mathbb{R}^{M \times 1}$

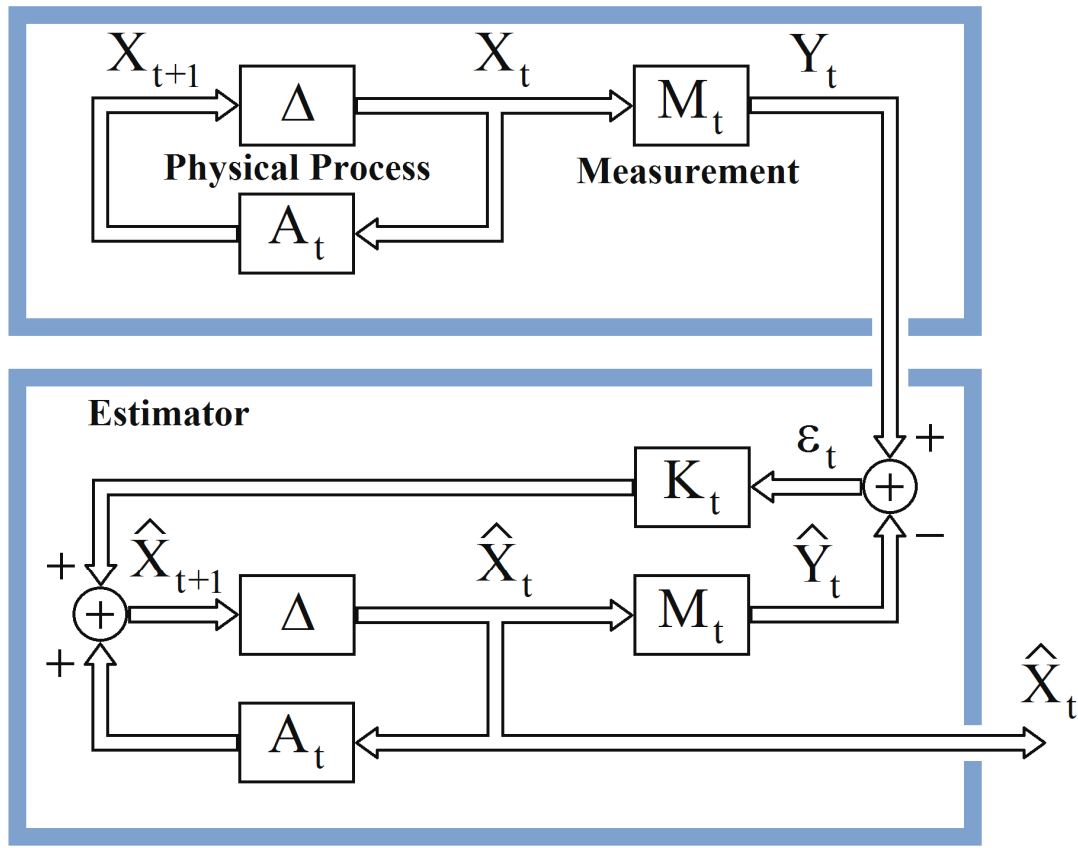

**Fig. 2** RLS.

*Original System:*

$$X_t = A_t X_{t-1} \tag{9}$$
$$Y_t = M_t X_t \tag{10}$$

*Estimator:*

$$\hat{X}_t = A_t \hat{X}_{t-1} + K_t \varepsilon_t \tag{11}$$
$$\hat{Y}_t = M_t \hat{X}_t \tag{12}$$

We can then define *a priori* and *a posteriori* (respectively) estimate error as:

$$\varepsilon_t^- = Y_t - \hat{Y}_t^- = Y_t - M_t \hat{X}_t^- \tag{13}$$

and

$$\varepsilon_t = Y_t - \hat{Y}_t = Y_t - M_t \hat{X}_t \tag{14}$$

The *a priori* estimate error covariance is then

$$\Xi\left\{\left(\varepsilon_t^-\right)\left(\varepsilon_t^-\right)^T\right\} = \Xi\left\{\left(Y_t - M_t \hat{X}_t^-\right)\left(Y_t - M_t \hat{X}_t^-\right)^T\right\} \tag{15}$$

where $\Xi\{\bullet\}$ means square error of "•", and $(\bullet)^T$ means transpose of "(•)".

On the other hand, the *a posteriori* estimate error covariance is

$$\Xi\left\{\varepsilon_t \varepsilon_t^T\right\} = \Xi\left\{\left(Y_t - M_t \hat{X}_t\right)\left(Y_t - M_t \hat{X}_t\right)^T\right\} \tag{16}$$

This adaptation process is based on the minimization of the mean square error criterion defined in the last equation. Developing Eq. 16, rearranging terms, and minimizing the mean square error with respect to $\hat{X}$, we obtain the Wiener filter to stationary signals, that is to say,

$$\hat{X} = R_{MM}^{-1} r_{MY} \tag{17}$$

where, $R_{MM}$ is the autocorrelation matrix **M** and $r_{MY}$ is the cross-correlation vector of **M** and **Y**. In the following, we formulate a recursive, time-update, adaptive formulation of Eq. 17. In fact, $R_{MM}$ can be expressed in recursive fashion as

$$R_{MM,t} = R_{MM,t-1} + M_t M_t^T \tag{18}$$

To introduce adaptability to the time variations of the signal statistics, the autocorrelation estimate in Eq. 18 can be windowed by an exponentially decaying window:

$$R_{MM,t} = \lambda R_{MM,t-1} + M_t M_t^T \tag{19}$$

where $\lambda$ is the so-called adaptation, or forgetting factor, and is in the range $0 < \lambda < 1$. Similarly, the cross-correlation vector can be calculated in recursive form as

$$r_{MY,t} = r_{MY,t-1} + M_t Y_t \tag{20}$$

Again this equation can be made adaptive using an exponentially decaying forgetting factor $\lambda$:

$$r_{MY,t} = \lambda r_{MY,t-1} + M_t Y_t \tag{21}$$

For a recursive solution of the least square error Eq. 21, we need to obtain a recursive time-update formula for the inverse matrix in the form

$$R_{MM,t}^{-1} = R_{MM,t-1}^{-1} + \text{Update}_t \tag{22}$$

where "Update$_t$" is an update factor to be actualized in each step time. After an extensive series of considerations, developments and replacements (such as $P_{MM,t} = R_{MM,t}^{-1}$), we get the following set of equations related to RLS adaptation algorithm [33-35] (very similar to Kalman's filter [28-32]).

Initial values:

- $P_{MM,0} = \delta I$ (being I the identity matrix and $\delta$ a number different to 0) (23)

- $\hat{X}_0 = \hat{X}_I$ (24)

Filter gain matrix:
$$K_t = P_{MM,t-1}^{-} M_t \left[ \lambda I + M_t^T P_{MM,t-1}^{-} M_t \right]^{-1} \tag{25}$$

Error signal equation:

$$\varepsilon_t^{-} = Y_t - M_t \hat{X}_{t-1}^{-} \tag{26}$$

Estimated states

$$\hat{X}_t = \hat{X}_{t-1}^{-} - K_t \varepsilon_t^{-} \tag{27}$$

Inverse correlation matrix update:

$$P_{MM,t} = \lambda^{-1} \left[ I - K_t M_t \right] P_{MM,t-1}^{-} \tag{28}$$

Discrete estimator time update equations

$$\hat{X}_t^{-} = A_t \hat{X}_{t-1} \tag{29}$$

$$P_{MM,t-1}^{-} = A_t P_{MM,t-1} A_t^T \tag{30}$$

Indeed, **A** and **M** are time-invariant [28-35]. In fact, we can dispense with the Eq. 30.

4.2 Quantum state estimator in noiseless environments

From Eq. 2, we have

$$|\psi\rangle_{pm} = |\varphi\rangle = \frac{\hat{M}_m |\psi\rangle}{\sqrt{\langle \psi | \hat{M}_m^\dagger \hat{M}_m | \psi \rangle}} \qquad (31)$$

being $\sqrt{\langle \psi | \hat{M}_m^\dagger \hat{M}_m | \psi \rangle}$ a norm of $\hat{M}_m$, as follow,

$$\|\hat{M}_m\| = \sqrt{\langle \psi | \hat{M}_m^\dagger \hat{M}_m | \psi \rangle} \qquad (32)$$

In fact, we can take any norm of $\hat{M}_m$, even for different $|\psi\rangle$ of the original. Thus,

$$|\varphi\rangle = \frac{\hat{M}_m}{\|\hat{M}_m\|} |\psi\rangle = \tilde{M}_m |\psi\rangle \qquad (33)$$

for each $m$, i.e., a battery of estimators, as show in Fig. 3.

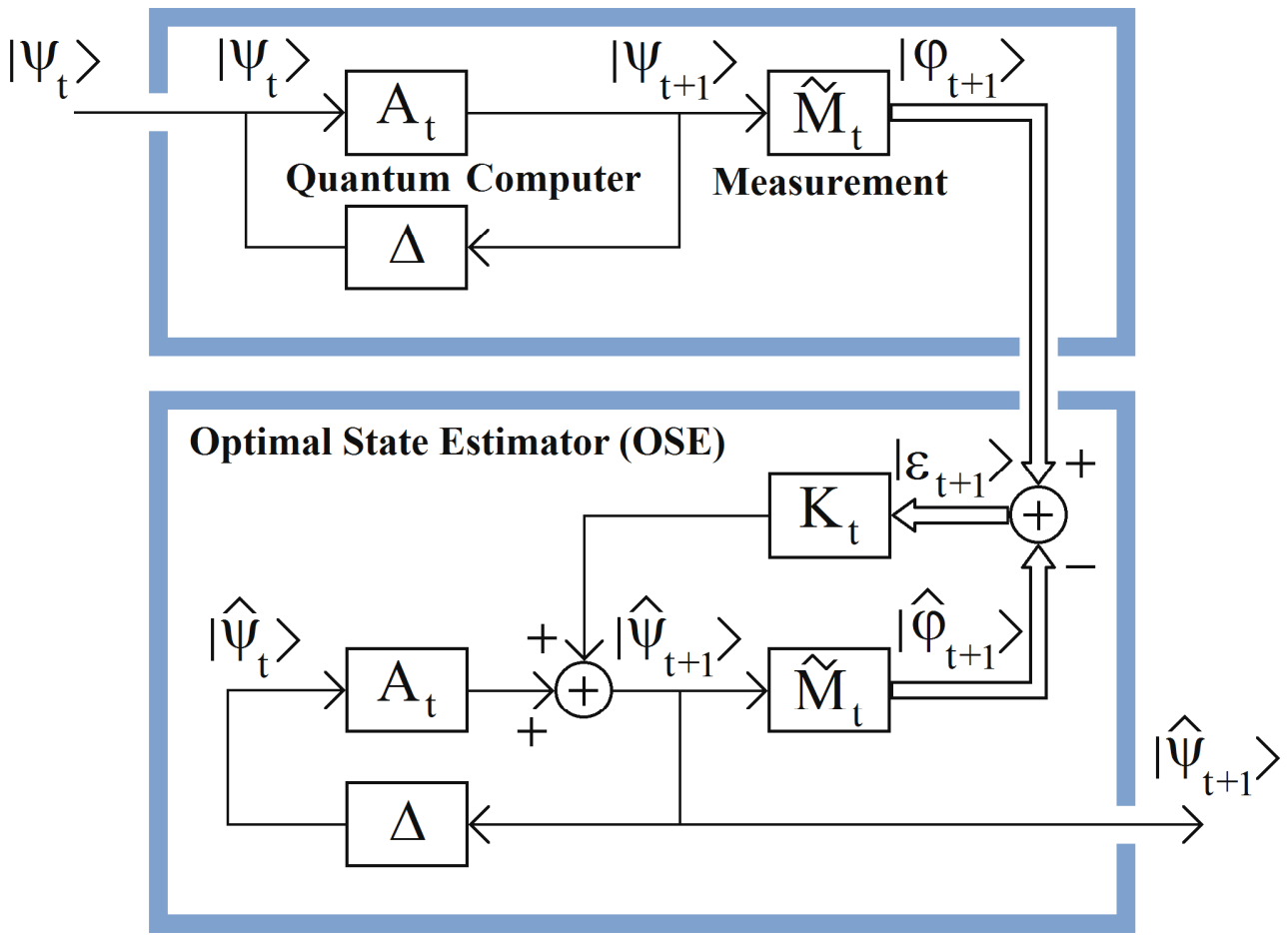

**Fig. 3** Modified RLS.

According Fig. 3, **A** will be the quantum algorithm (circuit or gate), and, we can get $|\psi\rangle$ for each $m$ with this estimator. Therefore, the complete set of equations is,

*Inside Quantum Computer:*

$$|\psi_{t+1}\rangle = A_t|\psi_t\rangle \quad \text{(quantum algorithm)} \tag{34}$$

$$|\varphi_{t+1}\rangle = \tilde{M}_t|\psi_{t+1}\rangle \quad \text{(quantum measurement)} \tag{35}$$

*Optimal State Estimator (OSE):*

$$|\hat{\psi}_{t+1}\rangle = A_t|\hat{\psi}_t\rangle + K_t|\varepsilon_{t+1}\rangle \tag{36}$$

$$|\hat{\varphi}_{t+1}\rangle = \tilde{M}_t|\hat{\psi}_{t+1}\rangle \tag{37}$$

*Estimation error:*

$$|\varepsilon_{t+1}\rangle = |\varphi_{t+1}\rangle - |\hat{\varphi}_{t+1}\rangle \tag{38}$$

Three important considerations:
- indeed, **A** is time-invariant, however, this metodology also resists the variant version (we can do similar considerations relating to **M**),
- really, OSE is a reorganized RLS/Kalman's filter, but it's the same algorithmically,
- we started with a poor measurement and evolution of OSE improves the accuracy of measurement

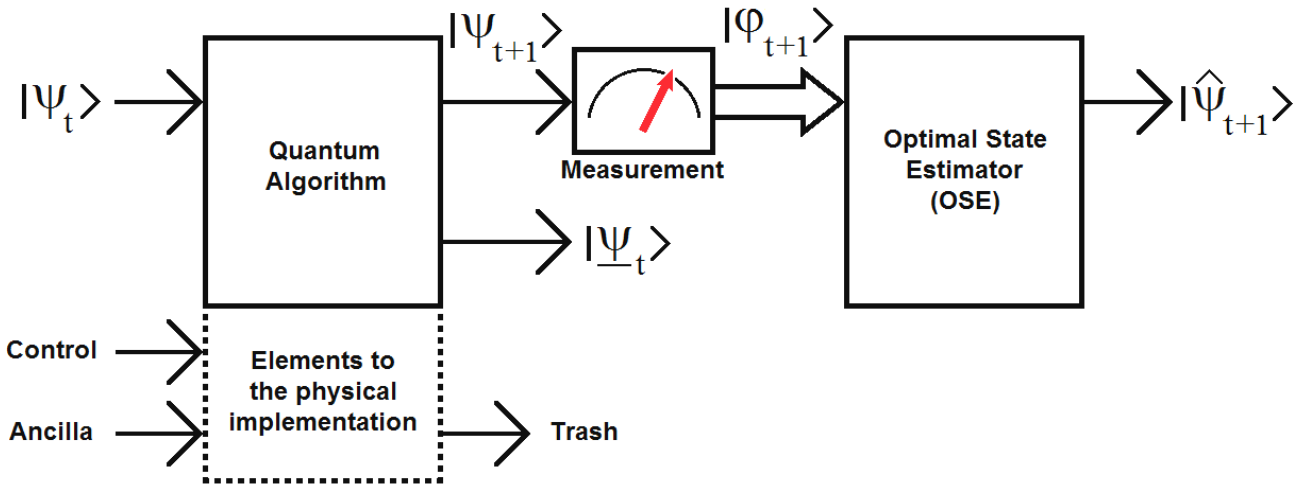

**Fig. 4** Quantum algorithm (circuit or gate), measurement and OSE.

Figure 4 shows the complete schematic of Fig. 1 but now with the OSE added to its output.

4.3 Quantum state estimator in noisy environments

We assume the existence of state and measurement noise, as seen in Fig. 5, with equation inside quantum computer

$$|\psi_{t+1}\rangle = A_t|\psi_t\rangle + N^s_{t+1} \quad \text{(quantum algorithm)} \tag{39}$$

$$|\varphi_{t+1}\rangle = \tilde{M}_t|\psi_{t+1}\rangle + N^m_{t+1} \quad \text{(quantum measurement)} \tag{40}$$

where, the random variables $N^s_{t+1}$ and $N^m_{t+1}$ represent the state and measurement noise, respectively. Both are assumed to be independent (of each other). In practice, the state noise covariance Q, and measurement noise

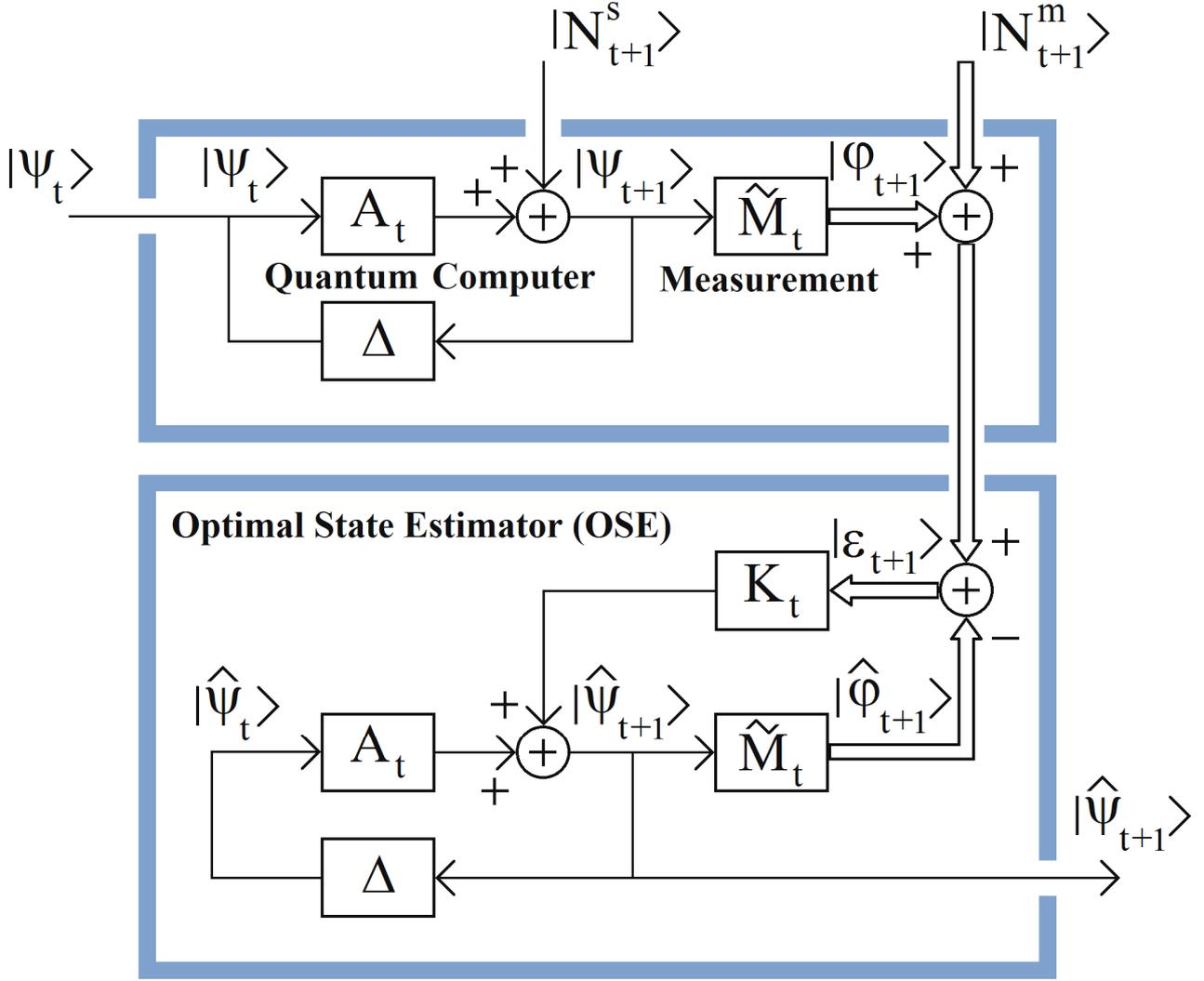

**Fig. 5** Modified Kalman's estimator for noisy environments.

covariance R matrices might change with each time step or measurement, however here we assume both are constant. Thus, only three equations change regarding to classic estimator, namely,

Filter gain matrix:

$$K_t = P^-_{MM,t-1} M_t \left[ R + M_t^T P^-_{MM,t-1} M_t \right]^{-1} \tag{41}$$

Inverse correlation matrix update:

$$P_{MM,t} = \left[ I - K_t M_t \right] P^-_{MM,t-1} \tag{42}$$

Discrete estimator time update equation

$$P^-_{MM,t-1} = A_t P_{MM,t-1} A_t^T + Q \tag{43}$$

However, and as the OSE is a linear system, we can move state noise to the output and work with a unique noise that represents both. Therefore, the last equation is not used.

All these noises may be associated with different factors: quantum noise [49, 50, 54-56], quantum decoherence [49, 57-62], and measurement errors [4-27]. The accuracy of our estimator (OSE) depends on two aspects:

- our ability to model these noises
- the greater or lesser presence of such noise in the experiment

## 5  Conclusions and Future Works

In this paper, we have presented an optimal estimator of quantum states based on a modified Kalman's Filter. Such estimator acts after state measurement, allowing obtain an optimal estimation of quantum state resulting to the output of any quantum algorithm (circuit or gate). Finally, the OSE allows us a complete estimation of the quantum state in a way quite more accurate than methods currently in use

**Acknowledgments** M. Mastriani thanks Sandra L. Rouget, CIO of DGIIT-ANSES, for her tremendous help and support.